# A SEARCH FOR NON-HIDDEN BROAD-LINE REGION SEYFERT 2 GALAXIES

GEORGI P. PETROV

*Department of Astronomy, Faculty of Physics,
St. Kliment Ohridski University of Sofia*

*Георги П. Петров.* ТЪРСЕНЕ НА СИЙФЪРТ 2 ГАЛАКТИКИ БЕЗ СКРИТА ОБЛАСТ НА ШИРОКИТЕ ЛИНИИ

Съгласно обединения модел на активни галактични ядра, Сийфърт 2 галактиките са физически еднакви със Сийфърт 1 обектите и притежават област на широките линии, но тя е скрита от наблюдателя, поради тяхната пространствена ориентация. През последните няколко години, различни автори докладват, че не всички Сийфърт 2 галактики притежават такава област. Събрали сме извадка от 38 Сийфърт 2 галактики за да намерим обекти без скрита област на широките линии. Използвайки теорията на Nicastro et al., която предполага съществуването на критична стойност на Едингтоновото отношение, под която стойност област на широките линии не може да се образува, ние намерихме 26 галактики без скрита област на широките линии. Също така открихме, че 5 от 26-те обекта може да са активни ядра от тип LINER (low-ionization nuclear emission-line region).

*Georgi P. Petrov.* A SEARCH FOR NON-HIDDEN BROAD-LINE REGION SEYFERT 2 GALAXIES

According to the unified model of active galactic nuclei, Seyfert 2 galaxies are physically the same as Seyfert 1 objects and they possess a broad-line region (BLR), but it is hidden from the observer due to their orientation. In the past few years, various authors reported that not all Seyfert 2 galaxies harbor a BLR. We compiled a sample of 38 Seyfert 2 galaxies to find non-hidden broad-line region (non-HBLR) objects. Using the theory of Nicastro et al. which suggests the existence of a critical value of the Eddington ratio below which BLR can't be formed, we found 26 non-HBLR Seyfert 2 candidates. We found also that 5 of these 26 non-HBLR objects could be low-ionization nuclear emission-line regions (LINERs).

*For contact:* Georgi P. Petrov, St. Kliment Ohridski University of Sofia, Faculty of Physics Department of Astronomy, 5 James Bourchier Blvd., 1164 Sofia, Bulgaria, Telephone: +359 2 81 61 201, Telefax: +359 2 962 52 76, E-mail: g_petrov@phys.uni-sofia.bg

## 1. INTRODUCTION

It has been widely accepted that every active galaxy possesses a central supermassive black hole (BH) with an accretion disk around it. This engine produces the observed hard X-ray continuum, which is strong enough to photoionize the broad-line region (BLR) near the source and the narrow-line region (NLR) located farther away from the nucleus.

Following the accepted Unified model for active galactic nuclei (AGNs) Seyfert 1 (Sy1) and Seyfert 2 (Sy2) galaxies are intrinsically the same and they differ only due to their orientation to the observer [1]. The main reason for this apparent difference is the presence of a dust torus around the nucleus and the BLR. When we observe Sy2 objects the dust torus is edge-on and hides the BLR from the observer, while the emission from the NLR is still visible. In the case of Sy1 galaxies the molecular torus is face-on and we have a direct view of the nuclear region. In support of the hypothesis of the Unified model Antonucci and Miller detected polarized broad emission lines (PBLs) in some Sy2 galaxies [2].

In the past few years, spectropolarimetric surveys have shown that there is a large fraction of Sy2 objects without PBLs, suggesting that not all Sy2 galaxies harbor a BLR [3, 4]. There are various explanations of the observed absence of a hidden broad-line region (HBLR) in some Sy2 galaxies. Lumsden and Alexander suggested that the detectability of the HBLR depends on the AGN luminosity [5]. Other authors proposed that the visibility of the HBLR is significantly determined by the nuclear obscuration [6].

Nicastro et al. found that the lack of BLR corresponds to low values of the accretion rate [7]. Their results show that all HBLR Sy2 galaxies have Eddington ratios above $10^{-3}$. The existence of this Eddington limit was confirmed also by other authors [8], although some of them found slightly different limit values [9].

In this paper we used the critical value of the Eddington ratio to find non-HBLR Sy2 candidates with physical absence of BLR.

## 2. DATA AND RESULTS

Ho et al. made a catalog of central stellar velocity dispersions (σ) of nearby galaxies [10]. 38 objects of the catalog are classified as Sy2 type by NASA/IPAC Extragalactic Database (NED)[1]. For these 38 Sy2s we used σ values adopted in the catalogue. The cosmological parameters assumed in this paper are $H_0 = 70$ km s$^{-1}$ Mpc$^{-1}$, $\Omega_\Lambda = 0.73$, and $\Omega_m = 0.27$.

---
[1] *http://ned.ipac.caltech.edu/*

We estimated the central BH masses ($M_{BH}$ [$M_\odot$]) of the 38 Sy2s using the empirical relation [11]:

$\log M_{BH} = 8.13 + 4.02 \log(\sigma / 200 \text{ km s}^{-1})$ .

At the same time, for each object of the sample we evaluated the Eddington ratio ($L_{Bol}/L_{Edd}$), where $L_{Bol}$ is the bolometric luminosity and the Eddington luminosity $L_{Edd}$ is given by [9]:

$L_{Edd} = 1.2 \times 10^{38} (M_{BH}/M_\odot)$ [erg s$^{-1}$].

In order to estimate $L_{Bol}$ for the objects from our sample, we took the observed $F_{[OIII]}^{obs}$ ($\lambda$ 5007) fluxes from References [4, 12, 13, 14, 15] and corrected them for the extinction by applying the relation [16]:

$$F_{[OIII]}^{cor} = F_{[OIII]}^{obs} \left[ \frac{(H\alpha/H\beta)_{obs}}{(H\alpha/H\beta)_0} \right]^{2.94},$$

where we adopted an intrinsic Balmer decrement $(H\alpha/H\beta)_0 = 3$. The observed Balmer decrement $(H\alpha/H\beta)_{obs}$ for each object was taken from the same references as the $F_{[OIII]}^{obs}$ fluxes.

The extinction corrected $L_{[OIII]}$ ($\lambda$ 5007) luminosities are given in Table 1. Following Lamastra et al. $C_{[OIII]} = L_{Bol} / L_{[OIII]}$, where $C_{[OIII]}$ is the bolometric correction. The mean value of $C_{[OIII]}$ in the luminosity ranges
$\log L_{[OIII]} = 38 - 40$, $40 - 42$ and $42 - 44$ is 87, 142 and 454, respectively [17].

The values of $L_{Bol}$ and ($L_{Bol}/L_{Edd}$) estimated from $L_{[OIII]}$ are listed in Table 1. In Fig.1 we have plotted ($L_{Bol}/L_{Edd}$) vs $M_{BH}$ for all 38 sample galaxies. The objects NGC 1068, NGC 2273 and NGC 4388 with observed HBLR in other surveys [4, 18] are shown as squares. The dashed line indicates the threshold of the Eddington ratio below which there is no Sy2s with HBLR. Nicastro et al. found this critical value ($L_{Bol}/L_{Edd}$) = $10^{-3}$ estimating $L_{Bol}$ from hard X-ray (2–10 keV) luminosity $L_X$ [7]. Recently, we obtained the same value, but estimating $L_{Bol}$ from $L_{[OIII]}$ [8] which is why we used this limit (the dashed line in Fig.1) to find non-HBLR Sy2 galaxies. We found 26 non-HBLR objects, all of them classified by NED as Sy2s. On the other hand, 5 of them occupy the area typically dominated by low-ionization nuclear emission-line regions (LINERs) – NGC 3607, NGC 4374, NGC 4378, NGC 4472 and NGC 4594. In Fig.1 this area is located in the lower right corner.

According to González-Martín et al. LINERs tend to have larger $M_{BH}$ than Sy2s [19]. Their results show that 84% of LINERs have $\log M_{BH} > 7.5$. At the same time LINERs have lower values of ($L_{Bol}/L_{Edd}$) than Sy2 galaxies. González-Martín et al. found for LINERs typical values of ($L_{Bol}/L_{Edd}$) ≈ $10^{-5}$, but they estimated $L_{Bol}$ from absorption corrected luminosity $L_X$. Transforming this value, in our case it should be ($L_{Bol}/L_{Edd}$) ≈ $10^{-6}$ and we assumed $10^{-5}$ as the upper limit of the Eddington ratio for the LINERs dominated area in Fig.1.

**Table 1.** Data for our sample of 38 Sy2 galaxies

| Name | z | $M_{BH}$ [$M_\odot$] | $L_{[OIII]}$ [$\frac{erg}{s}$] | From $L_{[OIII]}$ | | $L_X$ [$\frac{erg}{s}$] | From $L_X$ | | Ref. |
|---|---|---|---|---|---|---|---|---|---|
| | | | | $L_{Bol}$ [$\frac{erg}{s}$] | $\frac{L_{Bol}}{L_{Edd}}$ | | $L_{Bol}$ [$\frac{erg}{s}$] | $\frac{L_{Bol}}{L_{Edd}}$ | |
| NGC660  | 0.0028 | 7.35 | 40.11 | 42.26 | -3.17 | -     | -     | -     | [12] |
| NGC1058 | 0.0017 | 4.88 | 37.96 | 39.90 | -3.05 | 37.37 | 38.55 | -4.41 | [12, 14] |
| NGC1068 | 0.0038 | 7.77 | 42.89 | 45.55 | -0.29 | 42.95 | 44.42 | -1.42 | [4, 14] |
| NGC1358 | 0.0134 | 8.31 | 41.18 | 43.33 | -3.06 | 43.05 | 44.53 | -1.87 | [12, 9] |
| NGC1667 | 0.0152 | 7.83 | 42.35 | 45.00 | -0.91 | 42.31 | 43.79 | -2.13 | [4, 14] |
| NGC2273 | 0.0061 | 7.61 | 41.03 | 43.18 | -2.51 | 42.51 | 43.99 | -1.70 | [12, 14] |
| NGC2685 | 0.0029 | 6.81 | 38.92 | 40.86 | -4.03 | 39.72 | 40.90 | -3.99 | [12, 14] |
| NGC3079 | 0.0037 | 7.97 | 40.45 | 42.60 | -3.45 | 42.55 | 44.03 | -2.02 | [4, 14] |
| NGC3147 | 0.0093 | 8.29 | 40.82 | 42.97 | -3.40 | 41.87 | 43.35 | -3.03 | [15, 14] |
| NGC3185 | 0.0041 | 6.51 | 39.89 | 41.83 | -2.76 | 40.61 | 41.79 | -2.80 | [12, 14] |
| NGC3227 | 0.0039 | 7.46 | 40.45 | 42.60 | -2.94 | 41.55 | 43.03 | -2.51 | [12, 14] |
| NGC3254 | 0.0045 | 7.21 | 39.38 | 41.32 | -3.97 | -     | -     | -     | [12] |
| NGC3486 | 0.0023 | 6.17 | 38.29 | 40.22 | -4.02 | 39.10 | 40.27 | -3.97 | [12, 14] |
| NGC3607 | 0.0032 | 8.39 | 39.00 | 40.94 | -5.53 | 40.54 | 41.72 | -4.75 | [12, 19] |
| NGC3735 | 0.0090 | 7.51 | 40.77 | 42.92 | -2.67 | -     | -     | -     | [12] |
| NGC3941 | 0.0031 | 7.42 | 38.86 | 40.80 | -4.70 | 38.95 | 40.13 | -5.37 | [12, 14] |
| NGC3976 | 0.0083 | 8.04 | 39.56 | 41.50 | -4.61 | -     | -     | -     | [12] |
| NGC4169 | 0.0126 | 7.97 | 41.81 | 43.97 | -2.09 | -     | -     | -     | [12] |
| NGC4258 | 0.0015 | 7.60 | 39.06 | 41.00 | -4.69 | 40.76 | 41.93 | -3.75 | [12, 14] |
| NGC4303 | 0.0052 | 6.62 | 39.81 | 41.75 | -2.94 | -     | -     | -     | [12] |
| NGC4374 | 0.0035 | 8.88 | 38.92 | 40.86 | -6.10 | 41.31 | 42.79 | -4.18 | [12, 19] |
| NGC4378 | 0.0085 | 8.06 | 39.13 | 41.07 | -5.06 | -     | -     | -     | [12] |
| NGC4388 | 0.0084 | 6.77 | 41.65 | 43.80 | -1.05 | 42.39 | 43.87 | -0.98 | [4, 14] |
| NGC4472 | 0.0033 | 8.79 | 37.67 | 39.61 | -7.25 | 39.18 | 40.36 | -6.51 | [14] |
| NGC4477 | 0.0045 | 7.91 | 39.25 | 41.19 | -4.80 | 39.77 | 40.95 | -5.04 | [12, 14] |
| NGC4501 | 0.0076 | 7.81 | 39.87 | 41.81 | -4.08 | 40.17 | 41.34 | -4.55 | [12, 14] |
| NGC4565 | 0.0041 | 7.46 | 39.37 | 41.30 | -4.23 | 39.95 | 41.12 | -4.41 | [12, 14] |
| NGC4579 | 0.0051 | 7.79 | 39.70 | 41.64 | -4.23 | 41.25 | 42.73 | -3.14 | [13, 14] |
| NGC4594 | 0.0034 | 8.46 | 39.23 | 41.17 | -5.37 | 39.97 | 41.15 | -5.39 | [12, 19] |
| NGC4698 | 0.0034 | 7.61 | 38.78 | 40.72 | -4.97 | 39.03 | 40.20 | -5.49 | [12, 14] |
| NGC4725 | 0.0040 | 7.51 | 38.88 | 40.81 | -4.77 | 39.12 | 40.30 | -5.29 | [12, 14] |
| NGC4845 | 0.0041 | 7.43 | 39.65 | 41.59 | -3.92 | -     | -     | -     | [12] |
| NGC5194 | 0.0015 | 6.85 | 40.06 | 42.22 | -2.71 | 40.70 | 41.88 | -3.05 | [4, 14] |
| NGC5395 | 0.0117 | 7.57 | 39.36 | 41.30 | -4.35 | -     | -     | -     | [12] |
| NGC5631 | 0.0066 | 7.83 | 39.12 | 41.06 | -4.84 | -     | -     | -     | [12] |
| NGC5806 | 0.0045 | 7.31 | 38.45 | 40.39 | -4.99 | -     | -     | -     | [12] |
| NGC6951 | 0.0048 | 7.35 | 40.35 | 42.50 | -2.92 | -     | -     | -     | [12] |
| NGC7743 | 0.0057 | 6.72 | 40.42 | 42.57 | -2.23 | 41.47 | 42.95 | -1.85 | [12, 14] |

*Note:* Columns are: name of the galaxy, redshift z as reported in NED, logarithm of central BH mass, logarithm of extinction corrected [OIII] (λ 5007) luminosity, logarithm of bolometric luminosity and logarithm of Eddington ratio – predicted from $L_{[OIII]}$, logarithm of absorption corrected hard X-ray (2–10 keV) luminosity, logarithm of bolometric luminosity and logarithm of Eddington ratio – predicted from $L_X$, references for $L_{[OIII]}$ and $L_X$.

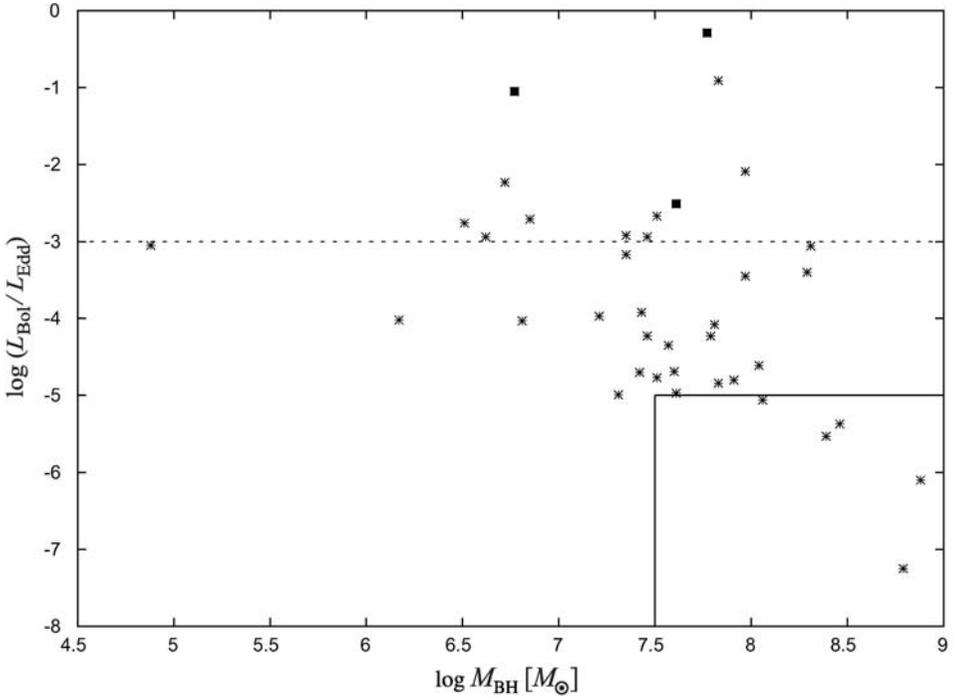

**Fig. 1.** ($L_{Bol}/L_{Edd}$) vs $M_{BH}$ diagram for 38 Sy2 galaxies. The Sy2 objects with observed HBLR are shown as squares and the rest Sy2s are shown as asterisks. The dashed line represents the threshold of the Eddington ratio below which there is no Sy2 with HBLR. At lower right corner is marked with solid lines the area dominated by LINERs (see the text).

We should mention that Marinucci et al. suggested a higher value of the Eddington limit [9]. Following their results the threshold between HBLR and non-HBLR Sy2s is log ($L_{Bol}/L_{Edd}$) = −1.9, where $L_{Bol}$ is derived from the absorption corrected hard X-ray (2–10 keV) luminosity $L_X$. Also, they proposed log $L_{Bol}$ = 43.90 as an additional separation value between the two types Sy2s.

In order to estimate $L_{Bol}$ and ($L_{Bol}/L_{Edd}$) from $L_X$, we took the absorption corrected $L_X$ luminosities from surveys [9, 14, 19], available for 26 objects of our sample (listed in Table 1). We used the relation $L_{Bol} = \kappa_{2-10\,keV} L_X$, where the hard X-ray bolometric corrections $\kappa_{2-10\,keV}$ are taken from Vasudevan and Fabian [20].

In Fig.2 we have plotted $L_{Bol}$ vs ($L_{Bol}/L_{Edd}$) for these 26 Sy2 galaxies. The dashed lines represent the separation values between HBLR and non-HBLR Sy2s given by Marinucci et al. [9]. As seen in the diagram, the Sy2 galaxies with observed HBLR are placed in the upper right corner, above the limits. On the other hand, the separation between objects which we previously suspected as HBLR and non-HBLR Sy2s is still visible at log ($L_{Bol}/L_{Edd}$) = −3.

The values of log $L_{Bol}$ for the two non-HBLR Sy2s for which this quantity exceeds 43.90 seem to be overestimated. We should note that $L_{Bol}$ depends on the $L_X$ absorption correction. Also, there is a separation between the objects that could be LINERs and the rest of the galaxies (see the upper left corner in Fig.2).

Generally, we prefer the value $10^{-3}$ of the Eddington ratio as a more reliable limit in our search for true non-HBLR Sy2s, especially when deriving ($L_{Bol}/L_{Edd}$) from $L_{[OIII]}$.

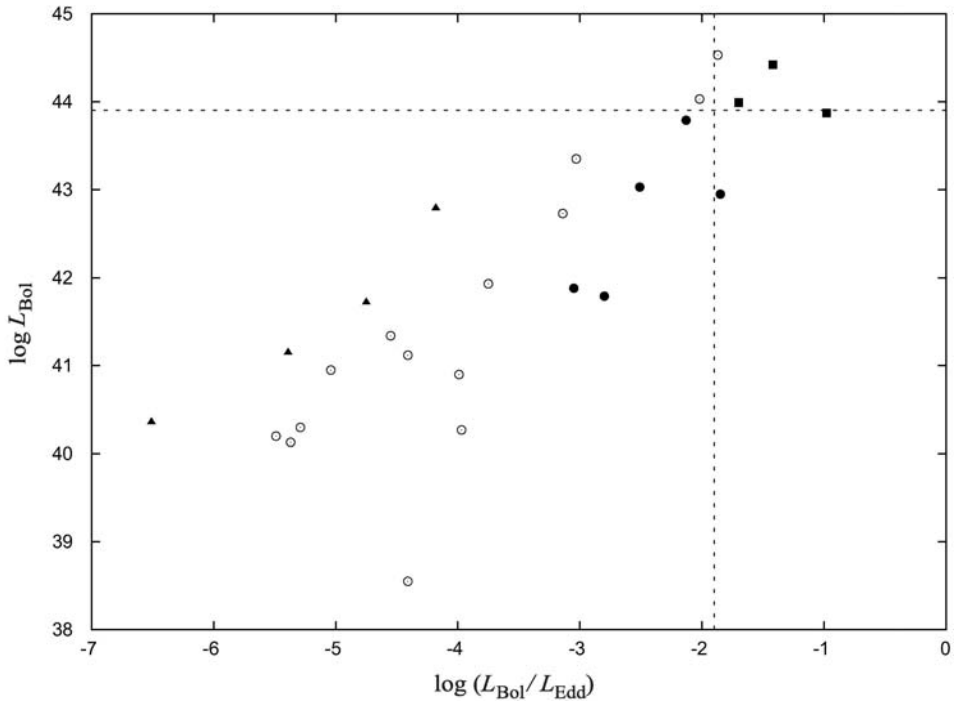

**Fig. 2.** $L_{Bol}$ vs ($L_{Bol}/L_{Edd}$) diagram for 26 Sy2 galaxies. The Sy2 galaxies with observed HBLR are shown as squares. The objects which we suspected as HBLR and non-HBLR Sy2s are shown as filled and open circles, respectively. The objects that could be LINERs are marked with triangles. The dashed lines represent the separation values between HBLR and non-HBLR Sy2s given by Marinucci et al. [9] (see the text).

## 3. CONCLUSIONS

Using the critical value of the Eddington ratio we found 26 non-HBLR candidates for true Sy2s with physical absence of BLR. We think that the limit value $(L_{Bol}/L_{Edd}) = 10^{-3}$ is reliable enough, when $L_{Bol}$ is estimated from $L_{[OIII]}$. The separation between these galaxies and the rest of the sample is still visible at $(L_{Bol}/L_{Edd}) = 10^{-3}$, also when $L_{Bol}$ is derived from $L_X$.

We found that 5 of these 26 non-HBLR objects could be LINERs. They occupy the LINERs dominated area in $(L_{Bol}/L_{Edd})$ vs $M_{BH}$ diagram (Fig.1) and also they are separated from the other Sy2s in $L_{Bol}$ vs $(L_{Bol}/L_{Edd})$ diagram (Fig.2).

**Acknowledgements.** I would like to thank to Lyubomir A. Ahtapodov for his helpful comments on the article.

This research has made use of the NASA/IPAC Extragalactic Database (NED) which is operated by the Jet Propulsion Laboratory, California Institute of Technology, under contract with the National Aeronautics and Space Administration.